\documentclass[amsmath,amssymb,12pt,superscriptaddress,nofootinbib]{revtex4}

\usepackage[dvipdfm]{graphicx}
\usepackage{bm}
\usepackage{color}

\newcommand{\dd}{\mathrm{d}}

\definecolor{DarkBlue}{rgb}{0,0,0.7} 

\definecolor{DarkRed}{rgb}{0.65,0,0} 

\definecolor{DarkGreen}{rgb}{0,0.6,0}

\begin{document}
\baselineskip5.5mm
\thispagestyle{empty}

{\baselineskip0pt
\leftline{\baselineskip14pt\sl\vbox to0pt{
               \hbox{\it Division of Particle and Astrophysical Science}
              \hbox{\it Nagoya University}
                             \vss}}
\rightline{\baselineskip16pt\rm\vbox to20pt{
\vss}}%
}

\author{Kengo Iwata}\email{iwata@gravity.phys.nagoya-u.ac.jp}
\affiliation{ 
Gravity and Particle Cosmology Group,
Division of Particle and Astrophysical Science,
Graduate School of Science, Nagoya University, 
Nagoya 464-8602, Japan
}

\author{Chul-Moon~Yoo}\email{yoo@gravity.phys.nagoya-u.ac.jp}
\affiliation{
Gravity and Particle Cosmology Group,
Division of Particle and Astrophysical Science,
Graduate School of Science, Nagoya University, 
Nagoya 464-8602, Japan
}

\vskip2cm
\title{Impact of small scale inhomogeneities \\
on observations of standard candles}

\begin{abstract}
\vskip0.5cm
We investigate the effect of small scale inhomogeneities 
on standard candle observations, such as type Ia supernovae (SNe) observations. 
Existence of the small scale inhomogeneities may cause  
a tension between SNe observations and other observations 
with larger diameter sources, such as 
the cosmic microwave background (CMB) observation. 
To clarify the impact of the small scale inhomogeneities, 
we use the Dyer-Roeder approach.
We determined the smoothness parameter $\alpha(z)$ 
as a function of the redshift $z$ 
so as to compensate the deviation of cosmological parameters for SNe  
from those for CMB. 
The range of the deviation which can be 
compensated by the smoothness parameter $\alpha(z)$ satisfying $0\leq\alpha(z)\leq1$
is reported. 
Our result suggests that the tension may give us the 
information of the small scale inhomogeneities through the smoothness parameter.

\end{abstract}

\maketitle
\pagebreak

\section{Introduction}

The modern cosmology has achieved great success 
based on the cosmological principle, which states that 
our universe is homogeneous and isotropic on 
large scales such as over 100Mpc. 
It is widely believed that 
the universe is well approximated by 
the Friedmann-Lema\^{\i}tre-Robertson-Walker (FLRW) model
over this scale. 
Cosmological parameters are estimated from 
observational data such as the type Ia supernovae (SNe) 
and the cosmic microwave background (CMB). 
Some tensions between independent observational data are often reported, 
e.g. the matter density parameter from SNe tends to be smaller than that from CMB. 
The tensions are more seriously considered than before 
due to recent progress of the observational accuracy. 
The reason for a tension would be a systematic error which originates from 
instruments or analysis, 
but it would be an unknown physical effect which may reveal the new physics and 
discovery. 
In this work, we propose use of such kinds of tension
to obtain information about small scale inhomogeneities.

As we mentioned, 
it has been reported that there is a tension between 
SNe observations and CMB observations \cite{Ade:2013zuv}.
This tension would originate from small scale inhomogeneities 
as is reported in Refs.~\cite{Fleury:2013uqa,Fleury:2013sna}. 
The real universe is inhomogeneous on the scales smaller than 100Mpc. 
It should be noted that 
the structure of the universe on the scales $\lesssim 10$kpc 
is unclear not only theoretically but also observationally. 
The theoretical difficulty mainly comes from nonlinear 
features of basic equations. 
Numerical simulations cost more to take the small scale inhomogeneities into account. 
On the other hand, 
observational investigation into much smaller scale inhomogeneities 
would be possible by using distant small diameter sources as probes. 
For instance, the typical diameter of SNe explosion is 
$10^{15}$cm. 
Therefore SNe can be regarded as point sources 
in cosmological observations.

In this paper, 
we focus on the comparison between 
SNe observations and 
other observations with much larger diameter sources, such as CMB observations. 
The observations of CMB is mainly affected by the large scale 
inhomogeneities and the influence from them can be estimated by using the perturbation theory. 
On the other hand, the observations of SNe 
may be significantly 
affected by the small scale inhomogeneities as well as the large scale ones. 
The effect of the inhomogeneities on the SNe observations is 
often estimated by using a ray shooting method 
with conventional models of inhomogeneity,
and 
it is suggested that the inhomogeneities do not 
significantly affect the cosmological parameter estimation. 
However, as is mentioned before, we 
do not have any reliable inhomogeneity model 
in the scale of the diameter of SNe, i.e. $10^{15}$cm. 
Therefore, in the precision cosmology era, 
it is important to estimate the impact of the 
inhomogeneities without any bias.

In oder to estimate the effect of the small scale inhomogeneities, as a first step, 
we use the simplest approximation method: the Dyer-Roeder (DR) approach \cite{DR1,DR2}. 
In this approach, we introduce a new parameter $\alpha$, 
which characterizes the small scale inhomogeneities. 
The parameter $\alpha$ is called the smoothness parameter and 
it is defined such that the fraction $\alpha$ of the matter is 
smoothly distributed, 
while the fraction $1-\alpha$ is bound in clumps. 
The light propagation is only affected by the smoothly 
distributed matter if we assume that the bundles of light 
lays propagate far from all the clumps. 
The validity of the DR approach is discussed in Ref.~\cite{Fleury:2014gha}. 
Observational constraints on the value of 
$\alpha$ is discussed in e.g. Ref.~\cite{BSL1} and references therein. 
In those analysis, $\alpha$ is assumed to be constant. 
However, the smoothness parameter does not necessarily have to be constant. 
The redshift dependence of the smoothness parameter was first discussed 
by Linder \cite{Linder1,Linder2}.

We consider the smoothness parameter 
is in the interval $0\leq\alpha(z)\leq1$
from its definition(but see Ref.\cite{Lima:2013rs} for $\alpha>1$ cases).  
The redshift dependence of the smoothness parameter 
should be related to the history of the structure formation of our universe. 
Naively thinking, 
the value of the smoothness parameter 
decreases with the progress of the structure formation. 
Therefore, 
the smoothness parameter is expected to be monotonically 
increasing function of the redshift.
In addition, since the early universe was totally homogeneous, 
it should asymptote to unity as $z\rightarrow\infty$.

Hereafter, we rely on the ``opacity hypothesis" proposed in Ref.~\cite{Fleury:2013sna}. 
This hypothesis states that all observed SNe have passed through 
the region far from clumps. 
It might be justified by the following reasons\cite{Fleury:2013sna}: 
the probability of passing through near clumps is too small, 
clumps may be bright enough to hide SNe behind it, 
strong gravitational lensing due to a clump makes it so bright 
that it is regarded as isolated exceptional source. 
Throughout this paper, we assume 
that the distance redshift relation of SNe 
is given by the DR distance with true cosmological parameters 
which describe global aspect of the universe.

In this paper, we demonstrate the impact of the inhomogeneities in the following way. 
We consider two sets of cosmological parameters which are taken 
from SNe observation ($\Omega^{\rm SNe},w^{\rm SNe}$) and another observation with larger 
source diameter, 
such as CMB anisotropy observation ($\Omega^{\rm CMB},w^{\rm CMB}$). 
We regard the cosmological parameters determined by 
the SNe observation are fictitious 
since the effect of the small scale inhomogeneities 
is not taken into account. 
Assuming the other set of cosmological parameters correctly describes 
the global aspect of our universe, 
we determine the smoothness parameter $\alpha(z)$ 
as a function of the redshift so that it compensates the deviation of 
($\Omega^{\rm SNe},w^{\rm SNe}$)
from ($\Omega^{\rm CMB},w^{\rm CMB}$). 
If the fixed smoothness parameter has the desired feature as a function of the redshift, 
it implies that the tension between ($\Omega^{\rm SNe},w^{\rm SNe}$) and 
($\Omega^{\rm CMB},w^{\rm CMB}$)
can be explained by the effect of the small scale inhomogeneities. 
At the same time, the tension gives us the information 
of the small scale inhomogeneities through the smoothness parameter.

This paper is organized as follows. 
In Sec.~\ref{sec:DR}, we briefly review the 
DR approach and derive the DR equation. 
How to determine the redshift dependent smoothness parameter 
is described in Sec.~\ref{sec:alpha}. 
In Sec.~\ref{sec.Results}, we show the behavior of 
the smoothness parameter as a function of the redshift, and 
Sec.~\ref{sec:S&C} is devoted to 
a summary.

In this paper, we use the geometrized units in which 
the speed of light and Newton's gravitational constant are one, 
respectively.

\section{Dyer-Roeder equation}
\label{sec:DR}

We assume that the universe is well described by the  Friedmann-Lema\^{\i}tre-Robertson-Walker (FLRW) model on the large scale. The Robertson-Walker metric is given by 
\begin{eqnarray}
 \dd s^2=\dd t^{2}-a^{2}(t)\left\{\frac{\dd r^{2}}{1-kr^2}+r^{2}(\dd \theta^{2}+\sin^2\theta \dd\phi^{2})\right\}  ,
\label{eq:metric}
\end{eqnarray}
where $a(t)$ is the scale factor, $k$ is the constant curvature. 
We consider non-relativistic matter and dark energy 
as energy components of the universe. 
The dark energy equation of state is given by 
$p_{\rm de}=w\rho_{\rm de}$, where $p_{\rm de}$ and $\rho_{\rm de}$ 
are the pressure and the energy density of the dark energy, respectively.
The energy-momentum tensor has 
the form of a perfect fluid: 
\begin{equation}
T^{\alpha\beta}=\left(\rho_{\rm m}+\rho_{\rm de}+p_{\rm de}\right)U^{\alpha}U^{\beta}-p_{\rm de}g^{\alpha\beta} ,
\label{eq:em tensor}
\end{equation}
where $\rho_{\rm m}$ is the energy density of the non-relativistic matter and 
$U^{\alpha}$ is the 4-velocity of a comoving volume element. 
From the Friedmann equation, the Hubble parameter is given by 
\begin{eqnarray}
H(z)&=&H_0\sqrt{\Omega_{\rm m}(1+z)^3+\Omega_{\rm de}(1+z)^{3(1+w)}+\Omega_{\rm k}(1+z)^2} \nonumber \\
&\equiv&H_0F(z;w,\Omega_{\rm m},\Omega_{\rm de},\Omega_{\rm k}) ,
\end{eqnarray}
where $H_0$ is the Hubble constant, $\Omega_{\rm m}$ and $\Omega_{\rm de}$ are 
the density parameter of the non-relativistic matter and the dark energy, 
respectively, and $\Omega_{\rm k}$ satisfies the equation 
\begin{equation}
{\Omega_{\rm m}}+{\Omega_{\rm de}}+{\Omega_{\rm k}}=1.
\label{eq:Friedmann0}
\end{equation}

The Dyer-Roeder equation is based on 
the Sachs optical equation\cite{Sachs}:
\begin{eqnarray}
\frac{\dd^2}{\dd v^2}\sqrt{A}&=&-\frac{1}{2}R_{\alpha\beta}k^{\alpha}k^{\beta}\sqrt{A}, 
\label{eq:A} 
\end{eqnarray}
where $v$ is the affine parameter, 
$A$ is the cross-sectional area of a light ray bundle, 
$R_{\alpha\beta}$ is the Ricci tensor and $k^{\alpha}$ is 
the null generator of the light rays. 
Here, we have neglected
the shear term (see e.g.\cite{1992grle.book.....S,sasaki}).
According to the field equation, 
we can replace the Ricci tensor by the energy-momentum tensor as follows:
\begin{eqnarray}
\frac{\dd^2}{\dd v^2}\sqrt{A}&=&-4\pi T_{\alpha\beta}k^{\alpha}k^{\beta}\sqrt{A} \nonumber \\
&=&-4\pi (\rho_{\rm m}+\rho_{\rm de}+p_{\rm de})(U_{\alpha}k^{\alpha})^2 \sqrt{A}.
\label{eq:A1} 
\end{eqnarray}	
Using the differential relation between the redshift and the affine parameter
\begin{equation}
\frac{\dd z}{\dd v}=(1+z)^2F(z;w,\Omega_{\rm m},\Omega_{\rm k}), 
\label{eq:rar}
\end{equation}
and the fact that the angular diameter distance 
is proportional to $\sqrt{A}$, we rewrite equation (\ref{eq:A}) as
\begin{align}
&F(z)\frac{\dd}{\dd z}\Bigl\{(1+z)^2F(z)\frac{\dd}{\dd z}{D}_{\rm A}(z)\Bigr\}
\cr
&\hspace{3cm}+\frac{3}{2}\big\{\Omega_{\rm m}(1+z)^3+\Omega_{\rm de}(1+w)(1+z)^{3(1+w)}\big\}{D}_{\rm A}(z)=0,
\label{eq:DR1}
\end{align}
where $\Omega_{\rm de}$ is given by $1-\Omega_{\rm m}-\Omega_{\rm k}$.
Since this equation depends on the parameters 
$w$, $\Omega_{\rm m}$ and $\Omega_{\rm k}$, 
we use the following expression: 
\begin{equation}
D_A(z)=D_A(z;w,\Omega_{\rm m},\Omega_{\rm k}). 
\end{equation}

In order to take account of small scale inhomogeneities, 
we introduce the smoothness parameter $\alpha(z)$ 
which describes the fraction of the smoothly distributed matter 
for each redshift. 
The fraction $1-\alpha(z)$ of the matter is clumped.
The case of $\alpha=1$ corresponds to a totally homogeneous universe, 
while for $\alpha=0$ all the matter is clumped. 
From the definition of the smoothness parameter, 
it is reasonable to consider only the interval 
\begin{equation}
0\leq\alpha(z)\leq1.
\label{eq:alpha cond}
\end{equation}
If a bundle of light rays passes through far away from the clumped regions, 
the light rays feel the gravitational field of the smoothly distributed matter. 
Therefore, we replace the  $\rho_{\rm m}$ in the energy-momentum tensor (\ref{eq:em tensor}) 
by $\alpha(z)\rho_{\rm m}$.
As a result, 
$\Omega_{\rm m}$ in the equation (\ref{eq:DR1}) is replaced 
by $\alpha(z)\Omega_{\rm m}$ 
and we get the Dyer-Roeder equation:
\begin{eqnarray}
&&F(z)\frac{\dd}{\dd z}\Bigl\{(1+z)^2F(z)\frac{\dd}{\dd z}{D}_{\rm DR}(z)\Bigr\}
\cr
&&\hspace{3cm}+\frac{3}{2}\big\{\alpha(z)\Omega_{\rm m}(1+z)^3+\Omega_{\rm de}(1+w)(1+z)^{3(1+w)}\big\}{D}_{\rm DR}(z)=0, 
\label{eq:DR}
\end{eqnarray}
where $D_{\rm DR}(z)$ is the Dyer-Roeder distance. 
Since the Dyer-Roeder distance depends on 
the cosmological parameters and the smoothness parameter, 
we use the following expression:
\begin{equation}
D_{\rm DR}(z)=D_{\rm DR}(z;w,\Omega_{\rm m},\Omega_{\rm k},\alpha(z)). 
\end{equation}
The boundary conditions for $D_{\rm A}$ and $D_{\rm DR}$ 
are given by 
\begin{eqnarray}
\begin{cases}
\left.D\right|_{z=0}=0, \\
\left.\frac{\dd D}{dz}\right|_{z=0}=1/H_0.
\end{cases}
\label{eq:initial}
\end{eqnarray}

\section{Determination of $\alpha(z)$}
\label{sec:alpha}

In this section, we explain 
the procedure to determine 
the smoothness parameter from observation.
As is mentioned in the introduction, 
we consider two sets of cosmological parameters 
($\Omega^{\rm SNe},w^{\rm SNe}$) and ($\Omega^{\rm CMB},w^{\rm CMB}$), 
which are taken from SNe observation and 
another observation with larger source diameter, 
such as CMB anisotropy observation, respectively.
Even though the observed luminosity distance for SNe has 
information of the small scale inhomogeneities,
the effect of the small scale inhomogeneities is not taken into account 
in the analysis of the observational data. 
Therefore, we consider that the effect of the small scale inhomogeneities 
may be an origin of the tension between ($\Omega^{\rm SNe},w^{\rm SNe}$) and 
($\Omega^{\rm CMB},w^{\rm CMB}$). 
Then, our question is that, {\em how large tension between 
($\Omega^{\rm SNe},w^{\rm SNe}$) and 
($\Omega^{\rm CMB},w^{\rm CMB}$) can be resolved by 
taking the small scale inhomogeneities into account?} 
Since, in the DR approach, 
the small scale inhomogeneities are
characterized by the smoothness parameter, 
we clarify how large tension between ($\Omega^{\rm SNe},w^{\rm SNe}$) and 
($\Omega^{\rm CMB},w^{\rm CMB}$) can be resolved by 
introducing the smoothness parameter $\alpha(z)$ 
which satisfies the condition \eqref{eq:alpha cond}.

First, we define the distance redshift relation $D^{\rm SNe}(z)$
based on the SNe observation as follows:
\begin{equation}
D^{\rm SNe}(z)=D_{\rm A}^{\rm SNe}(z;w^{\rm SNe},
\Omega^{\rm SNe}_{\rm m},\Omega^{\rm SNe}_{\rm k}), 
\label{eq:SNeFLRW}
\end{equation}
where $w^{\rm SNe}$, 
$\Omega^{\rm SNe}_{\rm m}$ and $\Omega^{\rm SNe}_{\rm k}$ 
are cosmological parameters given by the SNe observation. 
These cosmological parameters must be regarded as 
fictitious ones 
since we assume that the SNe observation is 
affected by the small scale inhomogeneities. 
Then, we assume the cosmological parameters 
$w^{\rm CMB}$, 
$\Omega^{\rm CMB}_{\rm m}$ and $\Omega^{\rm CMB}_{\rm k}$
precisely describe the global aspect of 
our universe. 
Under these assumptions and the DR approach, 
$D^{\rm SNe}(z)$ must be given by the DR distance with 
the true cosmological parameters
$w^{\rm CMB}$, 
$\Omega^{\rm CMB}_{\rm m}$ and $\Omega^{\rm CMB}_{\rm k}$. 
That is, 
\begin{equation}
D^{\rm SNe}(z)=D_{\rm DR}(z;w^{\rm CMB},\Omega^{\rm CMB}_{\rm m},\Omega^{\rm CMB}_{\rm k},\alpha(z)). 
\label{eq:SNeDR}
\end{equation}
Combining Eqs.~\eqref{eq:DR}, \eqref{eq:SNeFLRW} and \eqref{eq:SNeDR}, 
we obtain the smoothness parameter as follows:
\begin{eqnarray}
\alpha(z)&=&-\frac{2}{3\Omega_{\rm m}^{\rm{ CMB}}}
\biggl\{
\frac{F^{\rm CMB}(z)}{(1+z)^{3}D_{A}^{\rm SNe} (z)}
\frac{\dd}{\dd z}
\Bigl\{(1+z)^2F^{\rm CMB}(z)
\frac{\dd}{\dd z}D_{A}^{\rm SNe}(z)
\Bigr\}
\cr
&&\hspace{6cm}+\Omega_{\rm de}^{\rm CMB}(1+w^{\rm CMB})(1+z)^{3w^{\rm CMB}}
\biggr\},
\label{eq:alpha}
\end{eqnarray}
where $F^{\rm CMB}(z)=F(z;w^{\rm CMB},\Omega^{\rm CMB}_{\rm m},
\Omega^{\rm CMB}_{\rm k})$. 

The expression \eqref{eq:alpha} is singular at $z=0$ 
where $D_{A}^{\rm SNe} (z;w^{\rm SNe}, 
\Omega^{\rm SNe}_{\rm m},\Omega^{\rm SNe}_{\rm k})$ 
is zero. 
Expanding this expression in the vicinity of the center we find that 
the following condition must be satisfied to 
avoid singular behavior of the smoothness parameter:
\begin{equation}
3w^{\rm SNe}(\Omega_{\rm m}^{\rm SNe}+\Omega_{\rm k}^{\rm SNe}-1)+\Omega_{\rm k}^{\rm SNe}=3w^{\rm CMB}(\Omega_{\rm m}^{\rm CMB}+\Omega_{\rm k}^{\rm CMB}-1)+\Omega_{\rm k}^{\rm CMB}.
\label{eq:para cond}
\end{equation}
Therefore, once $w^{\rm CMB}$, $\Omega^{\rm CMB}_{\rm m}$ and 
$\Omega^{\rm CMB}_{\rm k}$ are fixed, 
we have two parameter degrees of freedom among 
$w^{\rm SNe}$, $\Omega^{\rm SNe}_{\rm m}$ and 
$\Omega^{\rm SNe}_{\rm k}$.

\section{Results}
\label{sec.Results}

Referring observational results in Ref.~\cite{Ade:2013zuv}, 
first, we fix the value of $\Omega^{\rm CMB}$ as 
follows:
\begin{equation}
\Omega_{\rm m}^{\rm CMB}=0.3175,~ \Omega_{\rm k}^{\rm CMB}=0.  
\end{equation}
In this paper, we consider the following 4 cases: 
\begin{itemize}
\item{(a): $w^{\rm SNe}=-1$}

\begin{enumerate}
\item{$w^{\rm CMB}=-1$ and one free parameter $\Omega_{\rm m}^{\rm SNe}$ 
($\Omega_{\rm k}^{\rm SNe}$ is determined by Eq.\eqref{eq:para cond}. )}

\item{$w^{\rm CMB}=-1.13$ and one free parameter $\Omega_{\rm m}^{\rm SNe}$ 
($\Omega_{\rm k}^{\rm SNe}$ is determined by Eq.\eqref{eq:para cond}. )}
\end{enumerate}

\item{(b): $\Omega_k^{\rm SNe}=0$}
\begin{enumerate}
\item{$w^{\rm CMB}=-1$ and one free parameter  $w^{\rm SNe}$
($\Omega_{\rm m}^{\rm SNe}$ is determined by Eq.\eqref{eq:para cond}. )}

\item{$w^{\rm CMB}=-1.13$ and one free parameter $w^{\rm SNe}$
($\Omega_{\rm m}^{\rm SNe}$ is determined by Eq.\eqref{eq:para cond}. )}
\end{enumerate}
\end{itemize}
In all the cases, 
$\Omega_{\rm de}^{\rm SNe}$ is determined by Eq.\eqref{eq:Friedmann0}.

\subsection*{(a): $w^{\rm SNe}=-1$}

Smoothness parameters for various $\Omega_{\rm m}^{\rm SNe}$ in the case of (a)-1 and 2
are shown in Figs.~\ref{fig:SN,lam-CMB,lam} and \ref{fig:SN,lam-CMB,w}, respectively.
\begin{figure}[htbp]
\begin{center}
\includegraphics[width=9cm]{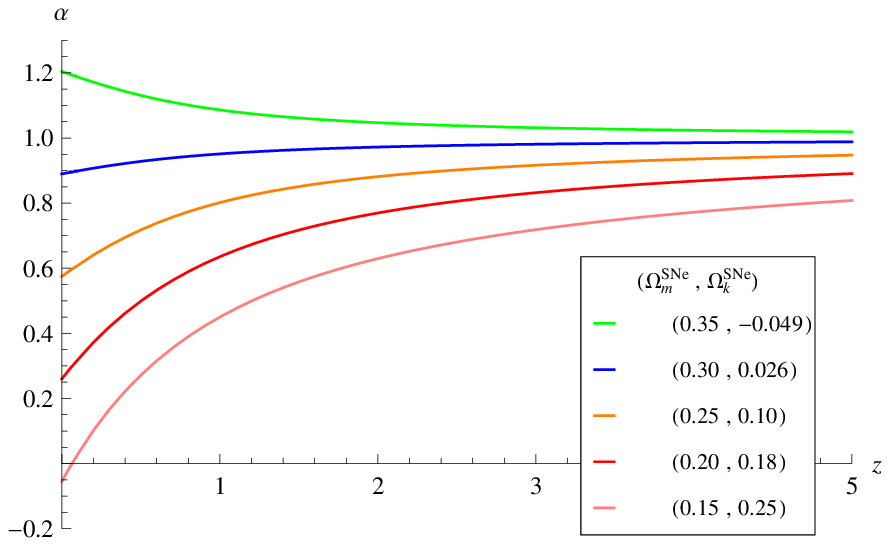}
\caption{Smoothness parameters as functions of $z$ for the case (a)-1. }
\label{fig:SN,lam-CMB,lam}
\end{center}
\end{figure}
\begin{figure}[htbp]
\begin{center}
\includegraphics[width=9cm]{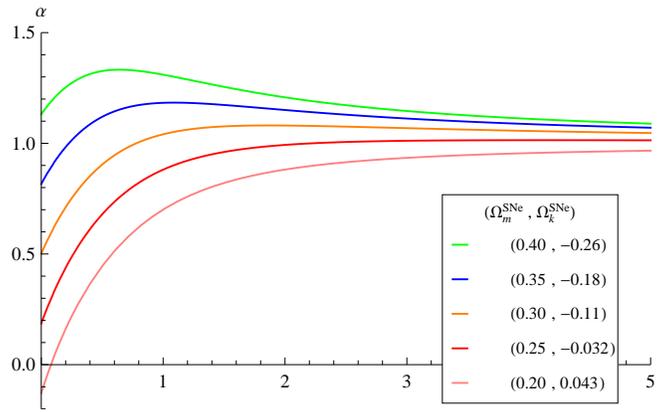}
\caption{Smoothness parameters as functions of $z$ for the case (a)-2. }
\label{fig:SN,lam-CMB,w}
\end{center}
\end{figure}

We require the smoothness parameter to be $0\leq\alpha(z)\leq1$.
In the case where this condition is not satisfied, 
we conclude that the tension between SNe and CMB observations 
cannot be resolved by only introducing the small scale inhomogeneities.
The smoothness parameter satisfies the condition 
\eqref{eq:alpha cond} in the following parameter region: 
\begin{equation}
\begin{array}{lll}
{\text (a)-1}&:&0.16\lesssim \Omega_{\rm m}^{\rm SNe}\leq0.3175, \\
{\text (a)-2}&:&0.22\lesssim \Omega_{\rm m}^{\rm SNe}\lesssim0.23. 
\end{array}
\label{(a)cond}
\end{equation}
The smoothness parameters are monotonically 
increasing functions of the redshift 
and it asymptotically approaches to unity 
if the matter density parameter 
is included in the parameter region \eqref{(a)cond}.
The parameter region given by \eqref{(a)cond} is consistent with
the fact that $\Omega^{\rm SNe}_{\rm m}$ tends to be 
smaller than $\Omega^{\rm CMB}_{\rm m}$\cite{Ade:2013zuv}.

\subsection*{(b): $\Omega_k^{\rm SNe}=0$}

Smoothness parameters for various $w^{\rm SNe}$ in the cases (b)-1 and 2
is shown in Figs.~\ref{fig:SN,w-CMB,lam} and \ref{fig:SN,w-CMB,w}, respectively.
\begin{figure}[htbp]
\begin{center}
\includegraphics[width=9cm]{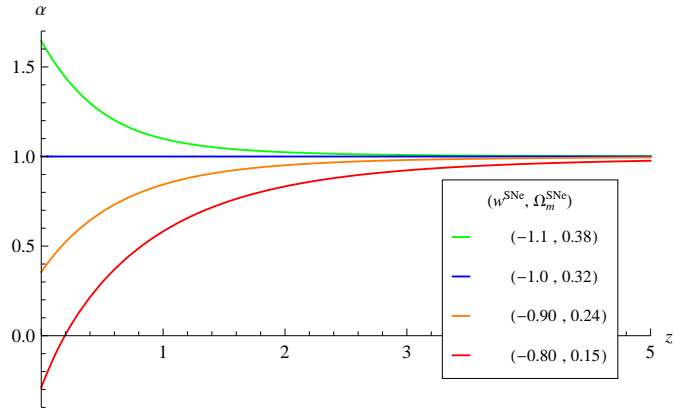}
\caption{Smoothness parameters as functions of $z$ for the case (b)-1. }
\label{fig:SN,w-CMB,lam}
\end{center}
\end{figure}
\begin{figure}[htbp]
\begin{center}
\includegraphics[width=9cm]{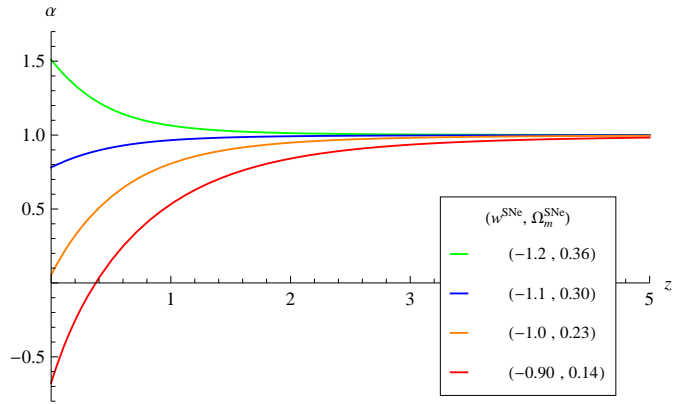}
\caption{Smoothness parameters as functions of $z$ for the case (b)-2. }
\label{fig:SN,w-CMB,w}
\end{center}
\end{figure}

The equation of state parameter for SNe $w^{\rm SNe}$ 
has to satisfy the following conditions to guarantee the condition $0\leq \alpha(z) \leq 1$. 
\begin{equation}
\begin{array}{lll}
{\text(b)-1}&:&-1.0\leq w^{\rm SNe}\lesssim-0.84, \\
{\text(b)-2}&:&-1.13\leq w^{\rm SNe}\lesssim-1.0. 
\end{array}
\label{(b)cond}
\end{equation}
The smoothness parameters monotonically increase with the redshift and 
asymptote to unity in the above parameter region.
In both cases, $w^{\rm SNe}$ must be larger than $w^{\rm CMB}$.
The results of the case (b) suggest that the equation of 
state parameters can differ from each other. 
From the result of the case (b)-1, 
even if dark energy is the cosmological constant, 
that is $w^{\rm CMB}=-1$, $w^{\rm SNe}$ can be different from unity 
by the effect of the small scale inhomogeneities.

\section{Summary and Conclusion}
\label{sec:S&C}

In this letter, 
we have proposed a way to obtain the information 
about the small scale inhomogeneities 
from the tension between SNe and other observations with 
much larger diameter sources, such as CMB. 
We have used the Dyer-Roeder approach 
to take account of the effect of the small scale inhomogeneities 
for distance-redshift relation of SNe.
The redshift dependent smoothness parameter $\alpha(z)$ 
has been introduced as 
the fraction of the smoothly distributed matter. 
Because of this definition, we required 
$\alpha(z)$ to be $0\leq \alpha(z) \leq1$.

We have determined the smoothness parameter  
as a function of the redshift 
so that it compensates the deviation of 
cosmological parameters estimated by SNe data
from those estimated by CMB data. 
The existence of such a $\alpha(z)$ satisfying $0\leq \alpha(z) \leq 1$ 
implies that the small scale inhomogeneities 
may cause the tension between CMB and SNe observations. 
That is, the small scale inhomogeneities may 
be an origin of significant systematic error in SNe observations
if we do not properly handle its effects.

We have introduced the smoothness parameter 
as a phenomenological parameter 
characterizes the small scale inhomogeneities. 
Therefore, the functional form of $\alpha(z)$ 
contains information about structure formation history of the universe. 
In this sence, we naturally expect that $\alpha(z)$ is 
monotonically increasing function of $z$ and 
asymptote to unity. 
We found that $\alpha(z)$ satisfies this property for our cases 
once we require $0\leq \alpha(z) \leq 1$. 
Our analysis implies that 
comparison between SNe and CMB observations 
may provide us the information about small scale inhomogeneities and 
its formation history.

It should be noted that, 
to make our proposal more realistic, 
several problems must be resolved. 
First, other origins of uncertainty in SNe observations, 
such as absorption effects and bias from 
poor understanding of the explorsion mechanism, 
must be resolved. 
Besides those effects, 
our analysis is based on the ``opacity hypothesis"\cite{Fleury:2013sna} 
which states that all observed SNe have path through 
the region far from clumps. 
Nevertheless, our proposal is unique one 
which has potential to probe 
extremely small scale ($>10^{15}$cm) inhomogeneities in cosmological observations 
in the future.

\end{document}